%% file: apssamp.tex
\documentclass[%
 amsmath,amssymb,
 aps, physrev,
 prapplied,
 reprint,
]{revtex4-2}

\usepackage{graphicx}
\usepackage{dcolumn}
\usepackage{bm}

\graphicspath{{figures/}}
\usepackage{amsmath}
\DeclareMathOperator*{\argmax}{\arg\!\max}

\usepackage{bbm}
\usepackage{xcolor}
\usepackage[normalem]{ulem}
\usepackage{gensymb}
\usepackage{upgreek}
\usepackage{enumitem}
\usepackage{hyperref}
\usepackage[capitalize]{cleveref}

\begin{document}

\preprint{APS/123-QED}

\title{\textbf{A Camera-Based Single-Volume, Monolithic Compton Camera} 
}%

\author{Alex Bocchieri}
 \email{Contact author: alexb826@gmail.com}
 \affiliation{University of Wisconsin - Madison, Madison, WI, USA
}
\author{Devanshi Jain}%
\affiliation{University of Wisconsin - Madison, Madison, WI, USA
}
\author{Andreas Velten}
\email{Contact author: velten@wisc.edu}
\affiliation{University of Wisconsin - Madison, Madison, WI, USA
}

\begin{abstract}
A Compton camera measures two events in a detector from individual gamma-rays to 
reconstruct the radiation source.
Conventional scintillator-based Compton cameras consist of two separate scintillator 
volumes or a single segmented scintillator, resulting in drawbacks regarding detector 
bulkiness and efficiency. 
Their scintillator readout techniques struggle to resolve two nearby events in one 
monolithic scintillator.
Single-photon avalanche diode (SPAD) cameras have recently been proposed for 
measuring scintillation events in various configurations.
SPAD array sensors enable high-resolution photon-counting acquisition at high rates 
that is easily scalable to millions of pixels. 
Prior work has explored different optics and measurement arrangements as well as 
demonstrated measurement of single interaction locations.
In this work, we extend these capabilities to demonstrate a practical radiation 
imaging device in the form of a Compton camera consisting of a SPAD camera and one 
compact, monolithic scintillator.
Our Compton camera prototype successfully reconstructs a gamma-ray source over 
multiple locations that span a 4$\pi$ steradian field of view.
Our approach is able to resolve two simultaneous events in one monolithic 
scintillator and enables Compton camera devices with compact, monolithic scintillator volumes for enhanced portability.

\end{abstract}

\maketitle

\input{sections/intro}

\input{sections/methods}

\input{sections/results}

\input{sections/discussion}

\section*{Data Availability}
The data that support the findings of this article and the code used to analyze the data may be obtained from the authors upon request.

\begin{acknowledgments}
This material is based upon work supported by the Department of Energy / National Nuclear Security Administration under Award Number(s) DE-NA0004196.
\end{acknowledgments}

\bibliography{apssamp}

\end{document}

%% file: sections/intro.tex
\section{Introduction}
The Compton camera is an electronic collimation method for reconstructing 
high-energy gamma-ray sources \cite{parajuli2022development,kim2024comprehensive,turecek2018compton,mihailescu2007speir}.
A Compton camera measures the 3D position and energy deposition of two events, namely Compton 
scattering and photoelectric absorption, from individual gamma-rays in a 
detector.
The radiation source is then reconstructed based on the kinematics of Compton scattering.
Compton cameras are applicable in nuclear security and nonproliferation \cite{vetter2018gamma,ziock2018principles,al2019passive,kataoka2013handy,mochizuki2017first},
medical imaging \cite{tashima2022compton}, 
and astronomy \cite{schonfelder1973telescope,schonfelder1984imaging,bloser2016balloon}.

Current scintillator-based Compton cameras predominantly use silicon photomultiplier 
(SiPM) arrays or similar sensors coupled to the scintillator's 
surface to measure double-events \cite{llosa2019sipm, parajuli2022development}.
A Compton camera must be able to spatially resolve and measure two simultaneous events.
Accordingly, Compton camera designs have consisted of a single-volume scintillator segmented with 
waveguides \cite{takahashi2013development, swanberg2018transparent, lee2020development, lee2022development, choi2025detection}
or two separate scintillator volumes \cite{lee2010compact,kataoka2013handy,kishimoto2017development,kim2018large,hosokoshi2019development, ogane2021simultaneous}.
However, inserting waveguides in a scintillator increases cost and decreases event 
detection efficiency.
Increasing distance between two separate scintillator volumes increases the overall 
space occupied by the detector and reduces sensitivity to wide-angle source locations and scatterings.
A single-volume, monolithic scintillator design could address both aforementioned designs' 
drawbacks, but current scintillation readout techniques for Compton cameras would struggle to 
resolve two nearby events in a monolithic scintillator due to overlapping light on the sensor.
High-resolution focal plane sensors combined with appropriate optics have been shown 
to effectively localize single point-like \cite{bocchieri2024scintillation,bocchieri2026kaleidoscopic,dai2026plenoptic} and track-like \cite{filipenko2014three, dieminger2026ultrafast, dieminger2026event} interactions within a monolithic scintillator, but not double-events.

We propose a camera-based scintillator readout technique to construct a single-volume, monolithic scintillator Compton camera. 
Specifically, we adopt a single-photon avalanche diode (SPAD) camera.
Prior simulations that extrapolated high photon counts have shown that double-events in one 
monolithic scintillator could be resolved with a SPAD camera, but experimental demonstrations were 
infeasible due to low light collection \cite{bocchieri2024scintillation}.
Now, we adopt a kaleidoscopic scintillator to obtain sufficient light collection and event 
spatial resolution to experimentally demonstrate a single-volume, monolithic scintillator Compton camera \cite{bocchieri2026kaleidoscopic}.
Furthermore, the SPAD camera's ability to resolve distinct nearby events allows for the miniaturization of 
scintillator volumes in Compton cameras.
Minimizing scintillator volume is important for reducing detector size and weight to 
enhance portability and integration into field-deployable devices.

The SPAD camera's sensor, the SPAD array, is fabricated with a highly scalable and integrated CMOS 
architecture that enables a high number of readout channels and high resolution for detecting 
single-photon arrivals that alternative sensors do not achieve.
Recently, there has been growing interest in adopting SPAD cameras 
\cite{bocchieri2024scintillation,bocchieri2026kaleidoscopic,dieminger2026ultrafast,dieminger2026event} 
and SPAD arrays \cite{franks2024demonstration,buhrer2025ai} 
for measuring scintillation events, driven largely by major advances in SPAD 
technology \cite{bruschini2019single}.
For example, SPAD arrays have achieved megapixel resolution, photon timetagging 
resolution of 100 ps, frame rates close to 100 kHz, and overall photon detection 
efficiencies approaching 70\% \cite{morimoto20213,ulku2018512}.
Previous works that adopt SPAD cameras have investigated individual 
components of radiation imaging, such as the camera lens or the localization of 
individual particle interactions, but demonstrations of 
actual radiation imaging capabilities are still lacking \cite{bocchieri2024scintillation,bocchieri2026kaleidoscopic,dieminger2026ultrafast,dai2026plenoptic, dieminger2026event}.

\begin{figure*}[ht!]
\centering
\includegraphics[width=\linewidth]{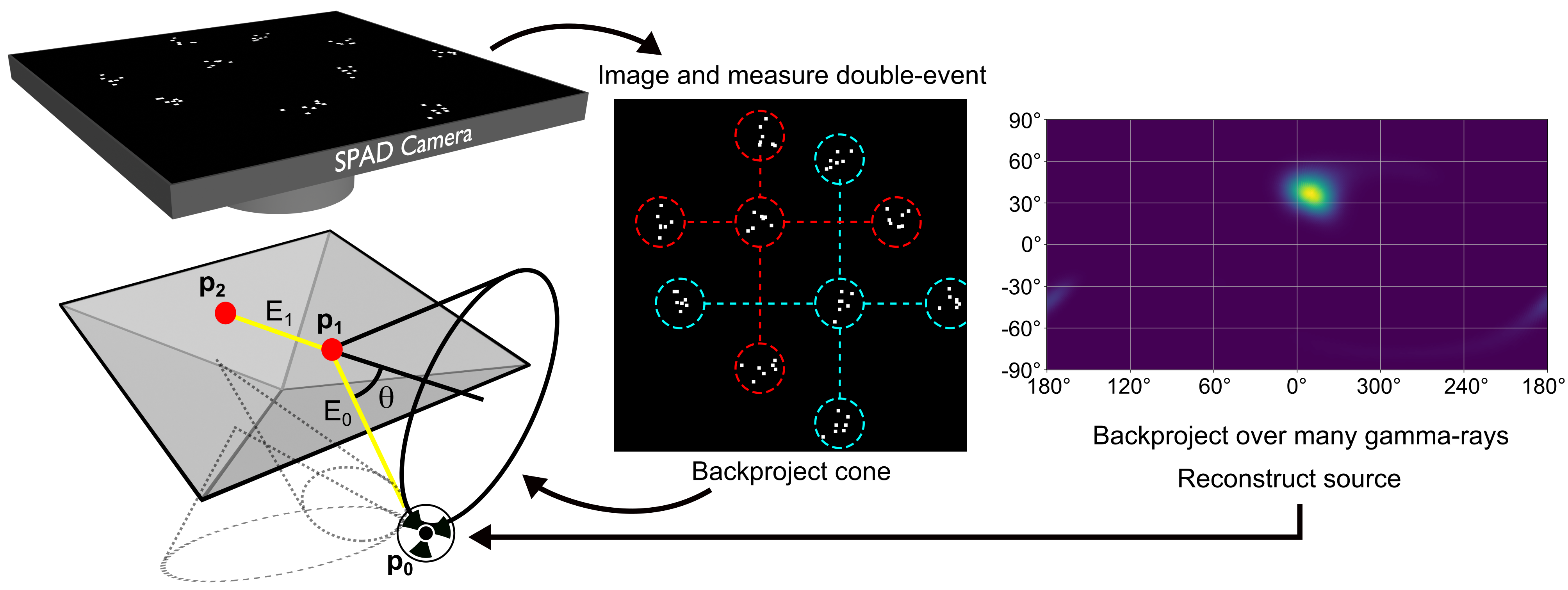}
\caption{Compton camera design and backprojection process.
A double-event in a kaleidoscopic scintillator from a gamma-ray is imaged with a 
SPAD camera and measured to backproject a cone.
The source's angular position is reconstructed by repeating this process over
many gamma-rays. 
} 
\label{fig:teaser}
\end{figure*}

We experimentally demonstrate our Compton camera prototype by reconstructing a 
gamma-ray source over multiple locations that span a 4$\pi$ steradian FOV.
We test our method with a compact scintillator volume, accepting 
reduced gamma-ray flux and lower event rates as a tradeoff.
Future optimization of the proposed design may enable a Compton camera with source 
reconstruction performance that is competitive with existing devices and provide 
benefits in terms of form factor and portability.
Our approach for resolving double-events in a monolithic scintillator is broadly applicable to any modality.
The design is illustrated in \cref{fig:teaser}.

Our contributions are as follows:
\begin{itemize}
\item A Compton camera that measures events with a camera and consists of one 
compact, monolithic scintillator.
\item An algorithm for measuring two events' locations and relative energy 
depositions in one image of a kaleidoscopic scintillator. 
\end{itemize}

%% file: sections/methods.tex
\section{Event Image Model}
\begin{figure}[h]
\centering
\includegraphics[width=\linewidth]{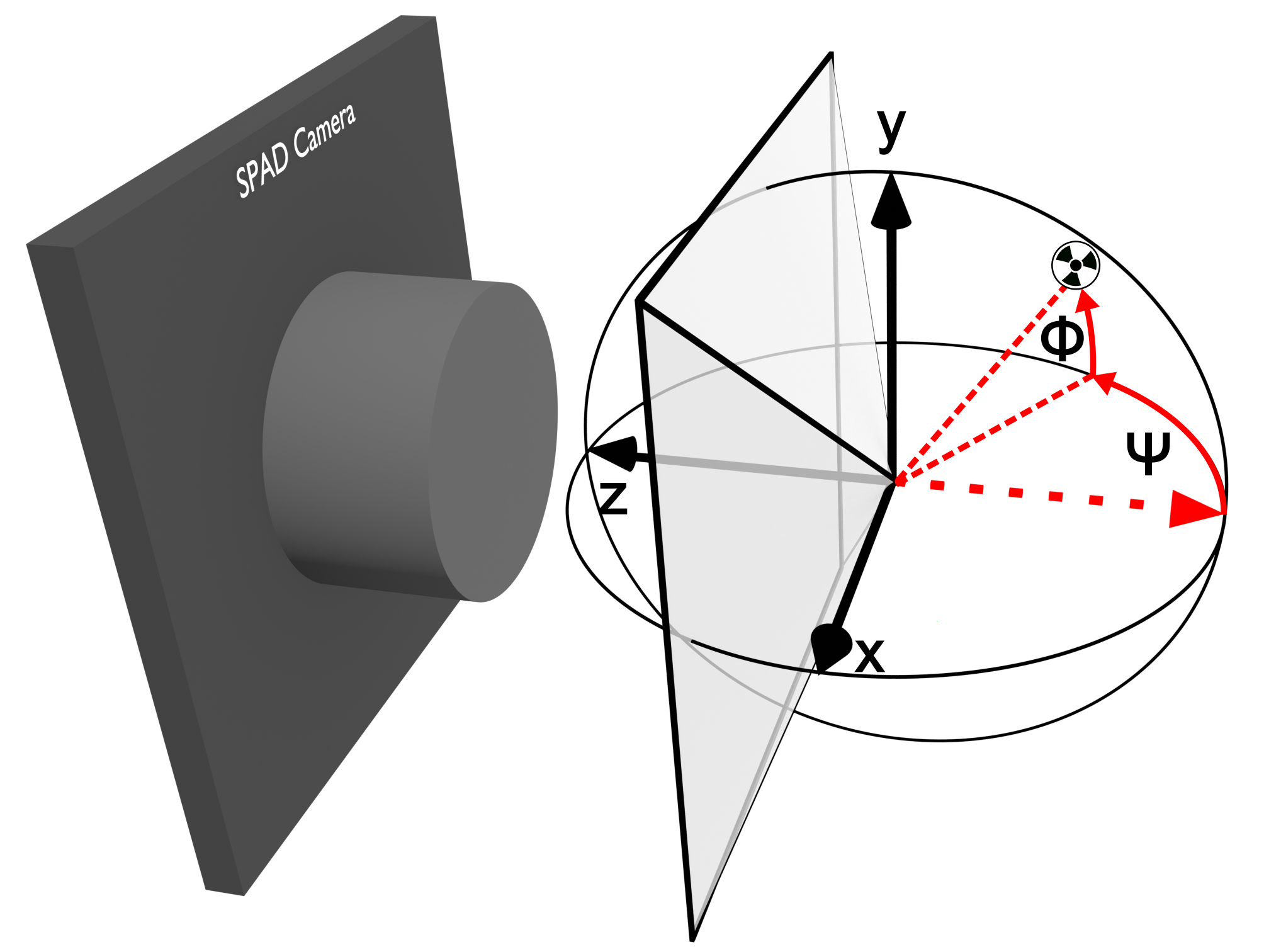}
\caption{Coordinate systems. The world coordinate system is denoted by the 
solid black arrows. The source is reconstructed over azimuth ($\Uppsi$) and 
elevation ($\Upphi$) denoted by the red arrows.} 
\label{fig:coord_system}
\end{figure}

We adopt a square pyramid geometry for the kaleidoscopic scintillator.
The world coordinate system origin is located at the scintillator's apex. 
The $z$-axis is directed toward the camera, and the $x$- and $y$-axes are 
perpendicular to the sensor's edges.
The camera's optical axis is along the $z$-axis.
The source reconstruction coordinate system is defined by azimuth ($\Uppsi$) and 
elevation ($\Upphi$) with respect to the origin at the scintillator apex.
The $(\Uppsi,\Upphi)=(0^\circ,0^\circ)$ direction is along the -$z$-axis. 
Azimuth is measured about the $y$-axis in the $xz$-plane, increasing 
counterclockwise when viewed from the +$y$ direction.
Elevation is measured relative to the $xz$-plane, with positive elevation in the 
+$y$ direction.
The source reconstruction map spans the full 4$\pi$ steradian FOV.
The coordinate systems are illustrated in \cref{fig:coord_system}.

An event and its mirror reflections are approximated as point sources of light whose 
locations are constrained according to the scintillator's geometry.
We model the camera with a thin lens and Gaussian point spread function.
Therefore, an event or mirror reflection's ($x$,$y$,$z$) location corresponds to a 
photon cluster in the image modeled as photon ($x$,$y$) coordinates distributed 
over an isotropic 2D Gaussian with a centroid $\bm{\mu}$ and standard deviation 
$\sigma$. 
The correspondence between an event or mirror reflection's ($x$,$y$,$z$) location 
and $\bm{\mu}$ and $\sigma$ is derived from depth from defocus and perspective 
projection and depends on the camera parameters. 
The equations defining this correspondence are provided in previous work \cite{bocchieri2026kaleidoscopic}.
The image of an event and its mirror reflections is modeled as a Gaussian mixture 
model (GMM), where all Gaussian components are parameterized by the event's 
($x$,$y$,$z$) location, as derived in previous work \cite{bocchieri2026kaleidoscopic}.
We assume each specular surface of the kaleidoscopic scintillator generates up to 
one mirror reflection for a total of up to four mirror reflections for one event.
Mirror reflections may be missing from the image due to finite mirrors, 
depending on their locations.

\section{Event Measurement Algorithm}
Our algorithm measures events in post-processing in two stages.
First, the algorithm detects whether an image contains two events.
Second, the algorithm measures the two events' 3D locations and 
energy depositions.
The double-event localization algorithm maximizes the GMM's likelihood 
using the EM algorithm.
The algorithm's double-event measurements are then used in the Compton camera 
backprojection method to reconstruct the gamma-ray source.

\subsection{Double-Event Detection} \label{sec:detection}
We use template matching to detect the presence of two events in an image.
One template consists of kernels that correspond to one event and its 
mirror reflections in an image.
We define kernel $k$, corresponding to event or mirror reflection $k$, over a square
region centered at $\bm{\mu_k}$ with side length 4$\sigma_k$.
Within this region, pixel intensities follow an isotropic 2D Gaussian distribution centered at $\bm{\mu_k}$ with standard deviation $\sigma_k$.
A template's pixels outside each kernel are zeroed.
$\bm{\mu_k}$ and $\sigma_k$ are computed based on the event location, 
scintillator geometry, and camera parameters.
$\sigma_k$ is assigned a minimum value of 10 pixels to prevent singularities.
We generate a set of templates for event locations over a 3D grid that spans 
the scintillator volume within the camera's FOV.
The grid's $x$-$y$ coordinates span the camera's lateral FOV. 
The grid's $z$-coordinates range from the scintillator's apex to the highest 
$z$-coordinate where four mirror reflections are observed in the camera's 
FOV.
This grid encompasses the measurable 3D region of the scintillator.

To detect two events in an image, we compute the Frobenius inner product between the 
image and each template and retain the template, $T_1$, that resulted in the 
maximum value as the first event's template.
Then, we remove all photons in the image that are within the kernels of $T_1$. 
We repeat this computation on the image with removed photons to obtain the second 
event's template, $T_2$.
We detect two events when, for both $T_1$ and $T_2$, the original image contains at 
least 3 kernels with at least 5 photons each, including a kernel corresponding to 
the event's photon cluster.
This provides initial estimates for event locations and classifies which mirror 
reflections are present in the image.
A mirror reflection is deemed to be present if its corresponding kernel contains at 
least 5 photons.
For $T_1$ and $T_2$, we record the number of kernels that contain at least 5 photons 
as $R_1$ and $R_2$ and the total number of photons within those kernels as $C_1$ and $C_2$, respectively.

\subsection{Double-Event Measurement}
To estimate the locations of two events in one image with $N$ photons and 
$K$ clusters, we optimize the following objective function:
\begin{equation} \label{eqn:opti_prob}
\argmax_{x_1,y_1,z_1,x_2,y_2,z_2, \bm{\pi}} Q
\end{equation}
where $Q$ is the expected log-likelihood of the complete-data likelihood function.

\begin{equation}
\begin{split}
Q = \sum_{i=1}^N \sum_{k=1}^K r_{ik} \bigg[ \text{log}(\pi_k) & + \text{log}(w_i) - \text{log}(2\pi{\sigma_k}^2) \\ & - \frac{w_i}{2{\sigma_k}^2}||\bm{t_i}-\bm{\mu_k}||_2^2 \bigg]
\end{split}
\end{equation}

\begin{align} \label{eqn:r_ik}
r_{ik} = \frac{\pi_k \mathcal{N}(\bm{t_i};\bm{\mu_k},{\sigma_k^2})}{\sum_{k'=1}^K \pi_{k'} \mathcal{N}(\bm{t_i};\bm{\mu_{k'}},{\sigma_{k'}^2})}
\end{align}
$(x_1,y_1,z_1)$ and $(x_2,y_2,z_2)$ are the two events' locations,
$\bm{t_i}$ is the $xy$-coordinate for photon $i$, 
$\bm{\pi}$ is a vector of mixing weights, $w_i$ is the weight applied to photon $i$, 
$\bm{\mu_k}$ and $\sigma_k$ are the parameters of Gaussian component $k$, 
and $r_{ik}$ is the posterior probability that photon $i$ belongs to cluster $k$.
$w_i = \sum_{j \in S_i^q} \text{exp} \left( -\frac{||\bm{t_i}-\bm{t_j}||_2^2}{\nu} \right)$,
where $S_i^q$ is the set of $q$ nearest neighbors of photon $i$, and $\nu$ is a 
positive scalar.
These terms are derived in previous work \cite{bocchieri2026kaleidoscopic}.

The initialized event locations and which $K$ clusters are present are obtained from the event detection stage.
Event location estimates are fine-tuned by optimizing \cref{eqn:opti_prob} 
with the EM algorithm.

We assume knowledge of the incident gamma-ray's energy and proportionally 
assign energy depositions to the two events based on their assigned photon counts. 
We assume all the gamma-ray's energy is deposited in the two events.
$\frac{5}{R_1}C_1$ and $\frac{5}{R_2}C_2$ counts are assigned to the two events, 
accounting for missing mirror reflections.

\section{Compton Camera Backprojection}
The Compton camera backprojection is based on the Compton scattering kinematics of 
an incident gamma-ray in the following process.
A gamma-ray with energy $E_0$ originating from a source located at $\bm{p_0}$ 
propagates into the scintillator and undergoes a Compton scattering event at $\bm{p_1}$.
During this event, the gamma-ray loses some energy, and its trajectory is 
deflected by an angle $\theta$ according to 
$\cos\theta = 1 - mc^2 (1/E_1 - 1/E_0)$. 
$E_1$ is the gamma-ray's energy after scattering, and $mc^2$ is the rest mass 
energy of an electron.
The gamma-ray continues propagating in the scintillator and undergoes a 
photoelectric absorption event at $\bm{p_2}$, during which it loses all of its 
energy and stops in its path.

The backprojection for one gamma-ray consists of a cone with vertex at $\bm{p_1}$, 
axis oriented along the vector from $\bm{p_2}$ to $\bm{p_1}$, and half-angle $\theta$.
The cone is constructed using measurements of the event pair's locations and 
energy depositions.
In simple backprojection, the source is reconstructed over many gamma-ray 
backprojections by accumulating the intersection of the cones' edges with the 
reconstruction surface. 
Cones are constructed using our algorithm's measurements of double-event locations 
and energy depositions.
The edges are assigned a width that adds blur to the reconstruction.
Source reconstruction can be deblurred and improved in quality using list-mode maximum likelihood expectation maximization (LM-MLEM) \cite{parra1998list, wilderman1998list}. 
The LM-MLEM reconstruction is defined as 
$\lambda_j^{n+1}=\frac{\lambda_j^n}{s_j}\sum\limits_i\frac{t_{ij}}{\sum_k t_{ik}\lambda_k^n}$
where $\lambda_j^n$ is the intensity of reconstruction pixel $j$ at the $n$th iteration, 
$s_j$ is the probability that a gamma-ray originating from reconstruction pixel $j$ is detected anywhere,
and $t_{ij}$ is the probability that a gamma-ray originating from reconstruction 
pixel $j$ results in the scatter-absorption pair measurement $i$.
The Compton scattering and backprojection processes are shown in \cref{fig:teaser}.

%% file: sections/results.tex
\section{Hardware Prototype}
\begin{figure}[t]
\centering
\includegraphics[width=\linewidth]{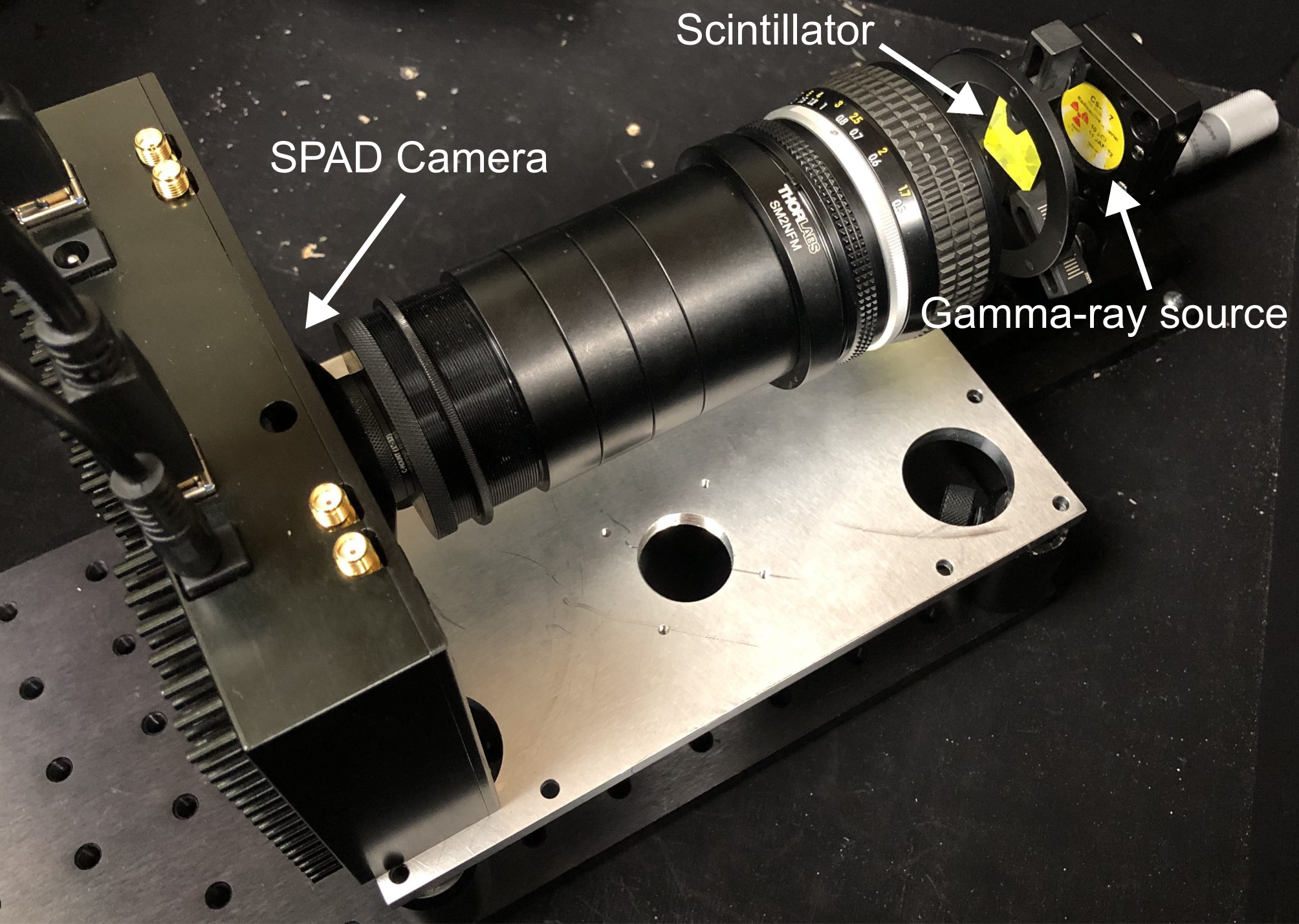}
\caption{Experimental setup.} 
\label{fig:exp_setup}
\end{figure}

We adopt the same kaleidoscopic scintillator and SPAD camera as in previous work \cite{bocchieri2026kaleidoscopic}.
The scintillator is a GAGG(Ce) crystal (Epic Crystal Co.) in the shape of a square 
pyramid with 5.77 mm height and a 20 mm base length.
All surfaces of the mirror except for the base surface are coated with enhanced 
specular reflector.
The SPAD camera consists of the SPAD512 array (Pi Imaging) with 512 $\times$ 512 
pixels and 16.4 $\mu$m pixel pitch and a 50 mm focal length lens (Nikkor) with f/1.2 aperture.
The lens and scintillator are separated by about 30 mm of air.
The SPAD camera is focused on the scintillator's apex, and its FOV spans 
5.5 $\times$ 5.5 mm.
The SPAD camera is oriented such that its optical axis is centered on the 
scintillator's apex and its sensor's edges are parallel with the scintillator's 
respective base surface edges.
After calibrating the camera parameters, the range in the $z$ dimension over which 
events can be measured is 3.5 mm.
This results in a measurable scintillator volume that is a square pyramid with 120 
degrees opening angle contained within a $5.5 \times 5.5 \times 3.5$ mm$^3$ 
rectangular prism for a total measurable volume of about 74 mm$^3$.
The setup is contained within a light-tight box.
The top 5\% of pixels with the highest dark count rates are zeroed.
The experimental setup is shown in \cref{fig:exp_setup}.

Data is collected by passively capturing 1-bit images with a 1.5 $\mu$s global 
shutter exposure time.
This exposure time is selected such that it is long enough to capture 
scintillation light emitted over scintillator's 150 ns decay constant and short 
enough to reduce dark count accumulation.
At this setting, the camera reads out images at about 19 kHz.
In post-processing, our algorithm analyzes all captured images and detects which 
images contain double-events for performing the Compton backprojection.

\section{Hardware Experiments}

\subsection{Algorithm Parameters}

The calibrated camera parameter values used in this work are $S_1=45$ mm, $S_2=64$ mm, 
$A=41.7$ mm, and $a=0.25$, where $S_1$ is the distance between the lens plane and 
sensor, $S_2$ is the distance between the lens plane and focal plane, $A$ is the 
lens diameter, and $a$ is a scaling factor relating the circle of confusion to
$\sigma_k$. 
These parameters relate an event's $(x,y,z)$ location to its corresponding photon 
cluster's $\bm{\mu}$ and $\sigma$, as defined in prior work \cite{bocchieri2026kaleidoscopic}.

We generate templates, as described in \cref{sec:detection}, over a grid that 
consists of 10 points equispaced in each dimension over $x \in [-2,2]$ mm, $y \in [-2,2]$ mm, and $z \in [0.1,3]$ mm.
A template is discarded if any of its kernels overlap or if its event's 
location is outside the scintillator.

In one M-step of the EM algorithm, we update $(x_1,y_1,z_1)$ and $(x_2,y_2,z_2)$ by 
running gradient ascent for 1,000 steps with a step size of 1e-7. 
All $\sigma_k$'s are clipped to a minimum value of 10 pixels.
The gradient of $\sigma_k$ is zeroed if $\sigma_k$ is clipped.
We assign weight $w_i$ to each photon using $q=10$ nearest neighbors and $\nu=10$ pixels.
The EM algorithm is terminated once the distance in estimated event locations 
between consecutive steps is less than 0.01 mm for both events.

Due to unknown ordering in scattering versus absorption events, 
we include both possible orderings in the list of scattering-absorption pair measurements.
We do not include an ordering that results in non-physical energy depositions 
according to Compton scattering kinematics.
The LM-MLEM $t_{ij}$ intensity is Gaussian centered on the backprojection's 
cone-sphere intersection pixels with standard deviation of 15 degrees.
The source is reconstructed over a sphere with a 20 mm radius and an average pixel 
separation of 0.9 degrees (Healpix nside=64).
$\lambda_j$ is initialized to all 1's.
$s_j$ is set to the uniform distribution.
LM-MLEM is run for 10 iterations.

\subsection{Source Reconstruction}
We evaluate our Compton camera prototype over a set of gamma-ray source 
locations using a 9 $\mu$Ci Cs-137 source (662 KeV; Spectrum Techniques). 
The source's radioactive material is approximately 3.1 mm in diameter.
We assign an uncertainty to the source's true location of $\pm$3 mm in each dimension.
The ambient temperature during capturing is 24 $^{\circ}$C,
and we obtain a median of 4 dark counts per image.
We capture a varying number of images over each source location.
We discard images with less than 60 counts.

The tested gamma-ray source locations and number of double-events detected are as follows:
\begin{enumerate}[label=(\alph*)]
\item
(0, 0, -20) mm, ($0^{\circ}$, $0^{\circ}$), 90 double-events
\item
(-20, 0, 0) mm, ($90^{\circ}$, $0^{\circ}$), 379 double-events
\item
(-20, 0, -20) mm, ($45^{\circ}$, $0^{\circ}$), 217 double-events
\item
(-10, 0, -20) mm, ($26.6^{\circ}$, $0^{\circ}$), 438 double-events
\item
(0, 20, -8) mm, ($0^{\circ}$, $68^{\circ}$), 250 double-events
\item
(0, 20, 15) mm, ($180^{\circ}$, $53.1^{\circ}$), 66 double-events
\item
(0, 10, -20) mm, ($0^{\circ}$, $26.5^{\circ}$), 140 double-events
\item
(20, 0, 0) mm, ($270^{\circ}$, $0^{\circ}$),  278 double-events
\item
(10, 0, -20) mm, ($333.4^{\circ}$, $0^{\circ}$),  546 double-events
\item
(0, -20, 0) mm, ($0^{\circ}$, $-90^{\circ}$),  326 double-events
\end{enumerate}

\begin{figure*}
\centering
\includegraphics[width=\linewidth]{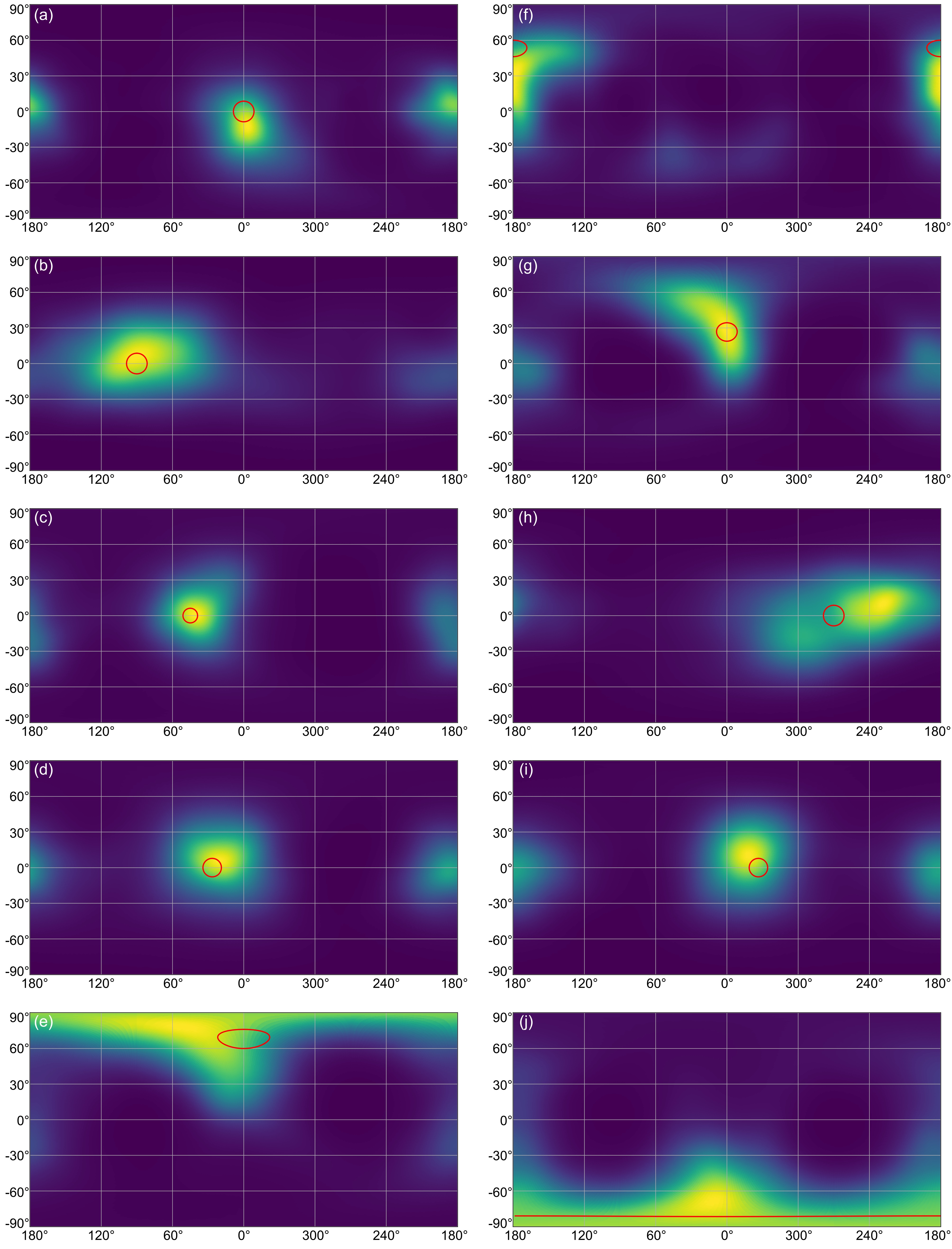}
\caption{Source reconstruction results. 
(a-j) The reconstruction map for each gamma-ray placement.
The red outline denotes the angle subtended by the gamma-ray source diameter 
and placement uncertainty. } 
\label{fig:recon_results}
\end{figure*}

\begin{figure*}
\centering
\includegraphics[width=\linewidth]{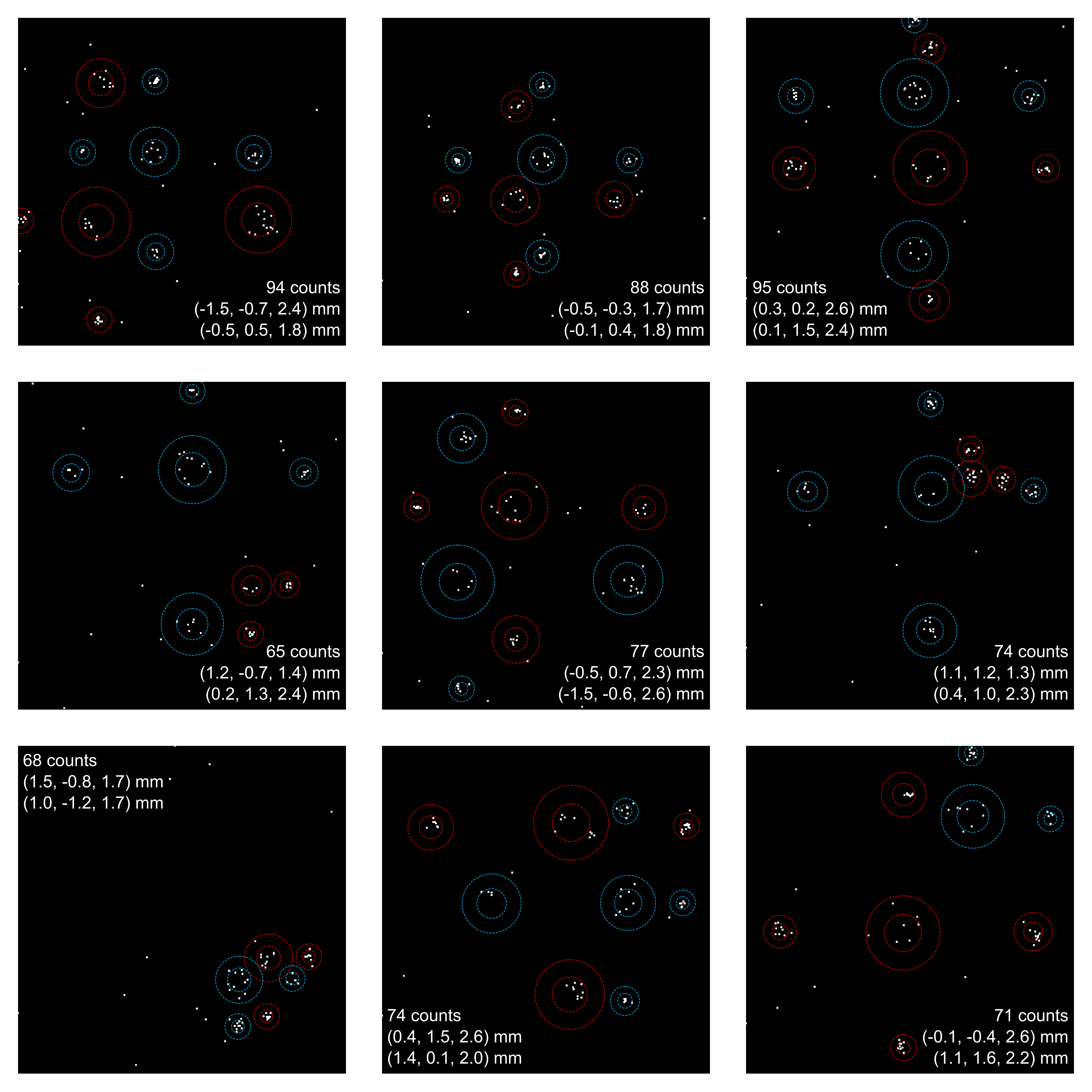}
\caption{Experimental images of double-events. 
The algorithm's estimated Gaussian components are illustrated as dashed circles centered on each Gaussian component's mean.
The inner and outer circles are one and two standard deviations in radius, respectively.
The circles' colors denote which event a Gaussian component belongs to.
Pixels with a photon are enlarged for visualization purposes.
The number of counts in the image and the algorithm's estimated event locations 
are shown in each image.
} 
\label{fig:exp_images}
\end{figure*}

The reconstructions after applying LM-MLEM are shown in \cref{fig:recon_results}.
Examples of experimental images containing two events and the algorithm's 
estimated event locations are shown in \cref{fig:exp_images}.
We observe a maximum of around 150 counts in an image.

%% file: sections/discussion.tex
\section{Discussion}
The reconstructions approximately match the source locations, as shown in  \cref{fig:recon_results}, 
and the GMM is correctly fit to the double-events in selected images shown in \cref{fig:exp_images}.
This demonstrates that our algorithm is able to measure two events in one image 
of a compact, monolithic scintillator, and that our prototype can successfully 
reconstruct sources over a 4$\pi$ steradian FOV.

The reconstruction results are noteworthy in regard to the scintillator's small form factor.
Our prototype's measurable scintillator volume of 74 mm$^3$ is much smaller than 
that of current Compton camera prototypes, which is typically on the order of 
$10 \times 10 \times 10$ mm$^3$ or higher \cite{kataoka2013handy,takahashi2013development, swanberg2018transparent, lee2020development, lee2022development, choi2025detection, lee2010compact,kishimoto2017development,kim2018large,hosokoshi2019development, ogane2021simultaneous}.
The small volume constrains scatter-absorption events to short separation distances,
causing event localization errors to have a greater impact on 
backprojection and making accurate source reconstruction more challenging.
In addition, the small volume causes most gamma-rays to pass through the 
scintillator without interacting, reducing event rates.
Despite these challenges, the prototype estimates event locations with sufficiently 
high spatial resolution to reconstruct the source. 

Source reconstruction results would likely improve if more events were captured.
Double-event counts are limited by long acquisition times due to the small 
scintillator volume, the low-activity source required by safety protocols, and
reducing individual image exposure.
Low source activity also forces the source to be placed close to the scintillator, 
making the source less point-like and increasing the uncertainty of its ground truth 
angular position.
Over all source locations, the algorithm detects an average of 173 double-events per 
1 billion captured images, or 1 double-event per 7 minutes of exposure.

There are several aspects of the prototype that future work may improve.
Photon counts per image are relatively low, even with adopting a kaleidoscopic 
scintillator, leading to relatively low energy resolution.
Future work could optimize the scintillator geometry by employing a smaller opening 
angle to image higher-order reflections and further increase light collection. 
The scintillator volume could be increased to capture more events or decreased for 
improved portability.
Modifying the camera optics would likely accompany such design choices.

The choice of SPAD camera could potentially be improved, specifically regarding 
photon detection efficiency and dark count rate.
We opted to reduce image exposure time to about 3\% of the image readout period to 
reduce dark count rates, resulting in longer acquisition times. 
The effect of dark count rates on the final measurement is complex and depends on 
the choice of algorithm and optical configuration among other things. 
Newer SPAD chips with lower dark count rates are becoming available, and dark count 
rate could be reduced by better active or passive cooling.
In general, a larger SPAD array area would reduce the demands on the optics and make 
it easier to achieve high energy resolution and position accuracy over larger volumes.

While this work develops a Compton camera, using a SPAD camera to image
double-events in a monolithic scintillator is broadly applicable to other radiation 
imaging modalities, including neutron scatter cameras \cite{mascarenhas2009results,braverman2018single,weinfurther2018model,manfredi2020single} 
and positron emission tomography (PET).
In PET, many monolithic scintillator designs assume one event location, 
whereas most annihilation photons interact two or more times \cite{gonzalez2021evolution}.
Our camera-based approach could spatially resolve and locate multiple 
events from one annihilation photon in a monolithic scintillator and potentially 
improve PET image reconstruction.

%% file: apssamp.bib
@PREAMBLE{
 "\providecommand{\noopsort}[1]{}" 
 # "\providecommand{\singleletter}[1]{#1}%" 
}

@article{bocchieri2024scintillation,
  title={Scintillation event imaging with a single photon avalanche diode camera},
  author={Bocchieri, Alex and Charbon, Edoardo and Velten, Andreas},
  journal={Communications Engineering},
  volume={3},
  number={1},
  pages={135},
  year={2024},
  publisher={Nature Publishing Group UK London}
}

@article{kataoka2013handy,
  title={Handy Compton camera using 3D position-sensitive scintillators coupled with large-area monolithic MPPC arrays},
  author={Kataoka, J and Kishimoto, A and Nishiyama, T and Fujita, T and Takeuchi, K and Kato, T and Nakamori, T and Ohsuka, S and Nakamura, S and Hirayanagi, M and others},
  journal={Nuclear Instruments and Methods in Physics Research Section A: Accelerators, Spectrometers, Detectors and Associated Equipment},
  volume={732},
  pages={403--407},
  year={2013},
  publisher={Elsevier}
}

@article{llosa2019sipm,
  title={SiPM-based Compton cameras},
  author={Llos{\'a}, G},
  journal={Nuclear Instruments and Methods in Physics Research Section A: Accelerators, Spectrometers, Detectors and Associated Equipment},
  volume={926},
  pages={148--152},
  year={2019},
  publisher={Elsevier}
}

@article{parajuli2022development,
  title={Development and applications of Compton camera—A review},
  author={Parajuli, Raj Kumar and Sakai, Makoto and Parajuli, Ramila and Tashiro, Mutsumi},
  journal={Sensors},
  volume={22},
  number={19},
  pages={7374},
  year={2022},
  publisher={MDPI}
}

@article{gonzalez2021evolution,
  title={Evolution of PET detectors and event positioning algorithms using monolithic scintillation crystals},
  author={Gonzalez-Montoro, Andrea and Gonzalez, Antonio J and Pourashraf, Shirin and Miyaoka, Robert S and Bruyndonckx, Peter and Chinn, Garry and Pierce, Larry A and Levin, Craig S},
  journal={IEEE Transactions on Radiation and Plasma Medical Sciences},
  volume={5},
  number={3},
  pages={282--305},
  year={2021},
  publisher={IEEE}
}

@article{vetter2018gamma,
  title={Gamma-Ray imaging for nuclear security and safety: Towards 3-D gamma-ray vision},
  author={Vetter, Kai and Barnowksi, Ross and Haefner, Andrew and Joshi, Tenzing HY and Pavlovsky, Ryan and Quiter, Brian J},
  journal={Nuclear Instruments and Methods in Physics Research Section A: Accelerators, Spectrometers, Detectors and Associated Equipment},
  volume={878},
  pages={159--168},
  year={2018},
  publisher={Elsevier}
}

@article{hosokoshi2019development,
  title={Development and performance verification of a 3-D position-sensitive Compton camera for imaging MeV gamma rays},
  author={Hosokoshi, Hiroki and Kataoka, Jun and Mochizuki, Saku and Yoneyama, Masaki and Ito, Soichiro and Kiji, Hiroaki and Nishi, Fumiya and Miyamoto, Shuji and Shima, Tatsushi},
  journal={Scientific reports},
  volume={9},
  number={1},
  pages={18551},
  year={2019},
  publisher={Nature Publishing Group UK London}
}

@article{kim2024comprehensive,
  title={A comprehensive review on Compton camera image reconstruction: From principles to AI innovations},
  author={Kim, Soo Mee and Lee, Jae Sung},
  journal={Biomedical Engineering Letters},
  volume={14},
  number={6},
  pages={1175--1193},
  year={2024},
  publisher={Springer}
}

@article{schonfelder1984imaging,
  title={The imaging Compton telescope COMPTEL on the gamma ray observatory},
  author={Schonfelder, V and Diehl, R and Lichti, GG and Steinle, H and Swanenburg, BN and Deerenberg, AJM and Aarts, H and Lockwood, J and Webber, W and Macri, J and others},
  journal={IEEE Transactions on Nuclear Science},
  volume={31},
  number={1},
  pages={766--770},
  year={1984},
  publisher={IEEE}
}

@article{lee2020development,
  title={Development of a position-sensitive 4$\pi$ Compton camera based on a single segmented scintillator},
  author={Lee, Hyounggun and Lee, Taewoong and Lee, Wonho},
  journal={IEEE Transactions on Nuclear Science},
  volume={67},
  number={12},
  pages={2511--2522},
  year={2020},
  publisher={IEEE}
}

@article{dieminger2026ultrafast,
  title={An ultrafast plenoptic-camera system for high-resolution 3D particle tracking in unsegmented scintillators},
  author={Dieminger, Till and Alonso-Monsalve, Sa{\'u}l and Alt, Christoph and Bruschini, Claudio and B{\"u}hrer, Noemi and Charbon, Edoardo and Kaneyasu, Kodai and Weber, Tim and Franks, Matthew and Sgalaberna, Davide},
  journal={Nature Communications},
  year={2026},
  publisher={Nature Publishing Group UK London}
}

@article{schonfelder1973telescope,
  title={A telescope for soft gamma ray astronomy},
  author={Sch{\"o}nfelder, Volker and Hirner, Andreas and Schneider, K},
  journal={Nuclear Instruments and Methods},
  volume={107},
  number={2},
  pages={385--394},
  year={1973},
  publisher={Elsevier}
}

@inproceedings{bocchieri2026kaleidoscopic,
    author={Bocchieri, Alex and Mamish, John and Appleyard, David and Velten, Andreas},
    title={Kaleidoscopic Scintillation Event Imaging},
    booktitle={Proceedings of the IEEE/CVF Conference on Computer Vision and Pattern Recognition (CVPR)},
    month={June},
    year={2026},
    pages={19749-19758}
}

@article{ulku2018512,
  title={A 512$\times$ 512 SPAD image sensor with integrated gating for widefield FLIM},
  author={Ulku, Arin Can and Bruschini, Claudio and Antolovi{\'c}, Ivan Michel and Kuo, Yung and Ankri, Rinat and Weiss, Shimon and Michalet, Xavier and Charbon, Edoardo},
  journal={IEEE Journal of Selected Topics in Quantum Electronics},
  volume={25},
  number={1},
  pages={1--12},
  year={2018},
  publisher={IEEE}
}

@inproceedings{morimoto20213,
  title={3.2 megapixel 3D-stacked charge focusing SPAD for low-light imaging and depth sensing},
  author={Morimoto, Kazuhiro and Iwata, J and Shinohara, M and Sekine, H and Abdelghafar, A and Tsuchiya, H and Kuroda, Y and Tojima, K and Endo, W and Maehashi, Y and others},
  booktitle={2021 IEEE International Electron Devices Meeting (IEDM)},
  pages={20--2},
  year={2021},
  organization={IEEE}
}

@article{tashima2022compton,
  title={Compton imaging for medical applications},
  author={Tashima, Hideaki and Yamaya, Taiga},
  journal={Radiological Physics and Technology},
  volume={15},
  number={3},
  pages={187--205},
  year={2022},
  publisher={Springer}
}

@article{mochizuki2017first,
  title={First demonstration of aerial gamma-ray imaging using drone for prompt radiation survey in Fukushima},
  author={Mochizuki, Saku and Kataoka, Jun and Tagawa, Leo and Iwamoto, Yasuhiro and Okochi, Hiroshi and Katsumi, Naoya and Kinno, Shuntaro and Arimoto, Makoto and Maruhashi, Takuya and Fujieda, Kazuya and others},
  journal={Journal of Instrumentation},
  volume={12},
  number={11},
  pages={P11014},
  year={2017},
  publisher={IOP Publishing}
}

@article{bloser2016balloon,
  title={Balloon flight test of a Compton telescope based on scintillators with silicon photomultiplier readouts},
  author={Bloser, PF and Legere, JS and Bancroft, CM and Ryan, JM and McConnell, ML},
  journal={Nuclear Instruments and Methods in Physics Research Section A: Accelerators, Spectrometers, Detectors and Associated Equipment},
  volume={812},
  pages={92--103},
  year={2016},
  publisher={Elsevier}
}

@article{lee2022development,
  title={Development of modified scintillator-based single-crystal position-sensitive 4$\pi$ Compton camera for a portable radiation imaging device},
  author={Lee, Hyounggun and Park, Jisu and Lee, Wonho},
  journal={Nuclear Instruments and Methods in Physics Research Section A: Accelerators, Spectrometers, Detectors and Associated Equipment},
  volume={1043},
  pages={167485},
  year={2022},
  publisher={Elsevier}
}

@article{choi2025detection,
  title={Detection efficiency and imaging performance of a 25.4$\times$ 25.4$\times$ 24 mm3 compton camera based on a single voxelized scintillator},
  author={Choi, Seonghee and Lee, Hyounggun and Lee, Wonho},
  journal={Radiation Physics and Chemistry},
  volume={233},
  pages={112754},
  year={2025},
  publisher={Elsevier}
}

@article{kim2018large,
  title={Large-area Compton camera for high-speed and 3-D imaging},
  author={Kim, Young-Su and Kim, Jae Hyeon and Lee, Junyoung and Kim, Chan Hyeong},
  journal={IEEE Transactions on Nuclear Science},
  volume={65},
  number={11},
  pages={2817--2822},
  year={2018},
  publisher={IEEE}
}

@article{kishimoto2017development,
  title={Development of a compact scintillator-based high-resolution Compton camera for molecular imaging},
  author={Kishimoto, A and Kataoka, J and Koide, A and Sueoka, K and Iwamoto, Y and Taya, T and Ohsuka, S},
  journal={Nuclear Instruments and Methods in Physics Research Section A: Accelerators, Spectrometers, Detectors and Associated Equipment},
  volume={845},
  pages={656--659},
  year={2017},
  publisher={Elsevier}
}

@article{swanberg2018transparent,
  title={Transparent ceramic garnet gamma-ray spectrometer with directionality},
  author={Swanberg, Erik L and Cherepy, Nerine J and Wihl, Brian M and Beck, Patrick R and Seeley, Zachary M and Hunter, Steven L and Fisher, Scott E and Payne, Stephen A and Kindem, Joel},
  journal={IEEE Transactions on Nuclear Science},
  volume={65},
  number={8},
  pages={2303--2309},
  year={2018},
  publisher={IEEE}
}

@article{parra1998list,
  title={List-mode likelihood: EM algorithm and image quality estimation demonstrated on 2-D PET},
  author={Parra, Lucas and Barrett, Harrison H},
  journal={IEEE transactions on medical imaging},
  volume={17},
  number={2},
  pages={228--235},
  year={1998},
  publisher={IEEE}
}

@inproceedings{wilderman1998list,
  title={List-mode maximum likelihood reconstruction of Compton scatter camera images in nuclear medicine},
  author={Wilderman, Scott J and Clinthorne, Neal H and Fessler, Jeffrey A and Rogers, W Les},
  booktitle={1998 IEEE Nuclear Science Symposium Conference Record. 1998 IEEE Nuclear Science Symposium and Medical Imaging Conference (Cat. No. 98CH36255)},
  volume={3},
  pages={1716--1720},
  year={1998},
  organization={IEEE}
}

@article{franks2024demonstration,
  title={Demonstration of particle tracking with scintillating fibres read out by a SPAD array sensor and application as a neutrino active target},
  author={Franks, Matthew and Dieminger, Till and Kaneyasu, Kodai and Sgalaberna, Davide and Bruschini, Claudio and Charbon, Edoardo and Kose, Umut and Li, Botao and Mos, Paul and Wayne, Michael and others},
  journal={The European Physical Journal C},
  volume={84},
  number={2},
  pages={202},
  year={2024},
  publisher={Springer}
}

@article{buhrer2025ai,
  title={AI-based particle track identification in scintillating fibres read out with imaging sensors},
  author={B{\"u}hrer, Noemi and Alonso-Monsalve, Sa{\'u}l and Franks, Matthew and Dieminger, Till and Sgalaberna, Davide},
  journal={Journal of Instrumentation},
  volume={20},
  number={05},
  pages={P05041},
  year={2025},
  publisher={IOP Publishing}
}

@article{filipenko2014three,
  title={Three-dimensional photograph of electron tracks through a plastic scintillator},
  author={Filipenko, Mykhaylo and Iskhakov, Timur and Hufschmidt, Patrick and Anton, Gisela and Campbell, Michael and Gleixner, Thomas and Leuchs, Gerd and Tick, Timo and Vallerga, John and Wagenpfeil, Michael and others},
  journal={The European Physical Journal C},
  volume={74},
  number={11},
  pages={3131},
  year={2014},
  publisher={Springer}
}

@article{turecek2018compton,
  title={Compton camera based on Timepix3 technology},
  author={Turecek, D and Jakubek, J and Trojanova, E and Sefc, L},
  journal={Journal of Instrumentation},
  volume={13},
  number={11},
  pages={C11022--C11022},
  year={2018}
}

@article{mihailescu2007speir,
  title={SPEIR: a Ge Compton camera},
  author={Mihailescu, L and Vetter, KM and Burks, MT and Hull, EL and Craig, WW},
  journal={Nuclear Instruments and Methods in Physics Research Section A: Accelerators, Spectrometers, Detectors and Associated Equipment},
  volume={570},
  number={1},
  pages={89--100},
  year={2007},
  publisher={Elsevier}
}

@article{ziock2018principles,
  title={Principles and applications of gamma-ray imaging for arms control},
  author={Ziock, KP},
  journal={Nuclear Instruments and Methods in Physics Research Section A: Accelerators, Spectrometers, Detectors and Associated Equipment},
  volume={878},
  pages={191--199},
  year={2018},
  publisher={Elsevier}
}

@article{mascarenhas2009results,
  title={Results with the neutron scatter camera},
  author={Mascarenhas, Nicholas and Brennan, James and Krenz, Kevin and Marleau, Peter and Mrowka, Stanley},
  journal={IEEE Transactions on Nuclear Science},
  volume={56},
  number={3},
  pages={1269--1273},
  year={2009},
  publisher={IEEE}
}

@article{al2019passive,
  title={Passive gamma-ray and neutron imaging systems for national security and nuclear non-proliferation in controlled and uncontrolled detection areas: Review of past and current status},
  author={Al Hamrashdi, Hajir and Monk, Stephen D and Cheneler, David},
  journal={Sensors},
  volume={19},
  number={11},
  pages={2638},
  year={2019},
  publisher={MDPI}
}

@article{bruschini2019single,
  title={Single-photon avalanche diode imagers in biophotonics: review and outlook},
  author={Bruschini, Claudio and Homulle, Harald and Antolovic, Ivan Michel and Burri, Samuel and Charbon, Edoardo},
  journal={Light: Science \& Applications},
  volume={8},
  number={1},
  pages={87},
  year={2019},
  publisher={Nature Publishing Group UK London}
}

@article{ogane2021simultaneous,
  title={Simultaneous measurements of single gamma ray of 131I and annihilation radiation of 18F with Compton PET hybrid camera},
  author={Ogane, Kenichiro and Uenomachi, Mizuki and Shimazoe, Kenji and Takahashi, Miwako and Takahashi, Hiroyuki and Seto, Yasuyuki and Momose, Toshimitsu},
  journal={Applied Radiation and Isotopes},
  volume={176},
  pages={109864},
  year={2021},
  publisher={Elsevier}
}

@inproceedings{takahashi2013development,
  title={Development of omnidirectional gamma-imager with stacked scintillators},
  author={Takahashi, Tone and Kawarabayashi, Jun and Tomita, Hideki and Iguchi, Tetsuo and Takada, Eiji},
  booktitle={2013 3rd International Conference on Advancements in Nuclear Instrumentation, Measurement Methods and their Applications (ANIMMA)},
  pages={1--4},
  year={2013},
  organization={IEEE}
}

@article{dai2026plenoptic,
  title={Plenoptic imaging of particle interactions in scintillation detectors},
  author={Dai, Xiang and Ho, Chi-Jui and Tandi, Kevin and Lee, Chang and Bocchieri, Alex and Parra, David and Peterson, Forrest and Sultan, Talha and Sutanto, Felicia and Velten, Andreas and others},
  journal={arXiv preprint arXiv:2607.01123},
  year={2026}
}

@article{lee2010compact,
  title={A compact Compton camera using scintillators for the investigation of nuclear materials},
  author={Lee, Wonho and Lee, Taewoong},
  journal={Nuclear Instruments and Methods in Physics Research Section A: Accelerators, Spectrometers, Detectors and Associated Equipment},
  volume={624},
  number={1},
  pages={118--124},
  year={2010},
  publisher={Elsevier}
}

@article{dieminger2026event,
  title={Event-by-event track imaging of charged particles with an ultrafast plenoptic camera},
  author={Dieminger, Till and Barbieri, Marco and Berardi, Vincenzo and Alonso-Monsalve, Sa{\'u}l and Kaneyasu, Kodai and Mos, Paul and Weber, Tim and W{\"u}thrich, Johannes and Bruschini, Claudio and Charbon, Edoardo and others},
  journal={arXiv preprint arXiv:2609.35509},
  year={2026}
}

@article{braverman2018single,
  title={Single-volume neutron scatter camera for high-efficiency neutron imaging and spectroscopy},
  author={Braverman, Joshua and Brennan, James and Brubaker, Erik and Cabrera-Palmer, Belkis and Czyz, Steven and Marleau, Peter and Mattingly, John and Nowack, Aaron and Steele, John and Sweany, Melinda and others},
  journal={arXiv preprint arXiv:1802.05261},
  year={2018}
}

@article{weinfurther2018model,
  title={Model-based design evaluation of a compact, high-efficiency neutron scatter camera},
  author={Weinfurther, Kyle and Mattingly, John and Brubaker, Erik and Steele, John},
  journal={Nuclear Instruments and Methods in Physics Research Section A: Accelerators, Spectrometers, Detectors and Associated Equipment},
  volume={883},
  pages={115--135},
  year={2018},
  publisher={Elsevier}
}

@inproceedings{manfredi2020single,
  title={The single-volume scatter camera},
  author={Manfredi, Juan J and Adamek, Evan and Brown, Joshua A and Brubaker, Erik and Cabrera-Palmer, Belkis and Cates, Joshua and Dorrill, Ryan and Druetzler, Andrew and Elam, Jeff and Feng, Patrick L and others},
  booktitle={Hard X-Ray, Gamma-Ray, and Neutron Detector Physics XXII},
  volume={11494},
  pages={121--131},
  year={2020},
  organization={SPIE}
}
